\documentclass[english,aps,preprint]{revtex4}
\usepackage[T1]{fontenc}
\usepackage[latin9]{inputenc}
\usepackage{float}
\usepackage{amsmath}
\usepackage{graphicx}

\makeatletter

\@ifundefined{textcolor}{}
{%
 \definecolor{BLACK}{gray}{0}
 \definecolor{WHITE}{gray}{1}
 \definecolor{RED}{rgb}{1,0,0}
 \definecolor{GREEN}{rgb}{0,1,0}
 \definecolor{BLUE}{rgb}{0,0,1}
 \definecolor{CYAN}{cmyk}{1,0,0,0}
 \definecolor{MAGENTA}{cmyk}{0,1,0,0}
 \definecolor{YELLOW}{cmyk}{0,0,1,0}
}

\@ifundefined{showcaptionsetup}{}{%
 \PassOptionsToPackage{caption=false}{subfig}}
\usepackage{subfig}
\makeatother

\usepackage{babel}
\begin{document}

\title{``Inchworm Filaments'': Motility and Pattern Formation}

\author{Nash Rochman}

\email{nashdeltarochman@gmail.com}

\affiliation{Department of Chemical and Biomolecular Engineering, The Johns Hopkins
University }

\author{Sean X. Sun}

\email{ssun@jhu.edu}

\affiliation{Departments of Mechanical Engineering and Biomedical Engineering,\\
The Johns Hopkins University, Baltimore MD 21218}
\begin{abstract}
In a previous paper, we examined a class of possible conformations
for helically patterned filaments in contact with a bonding surface.
In particular, we investigated geometries where contact between the
pattern and the surface was improved through a periodic twisting and
lifting of the filament. A consequence of this lifting is that the
total length of the filament projected onto the surface decreases
after bonding. When the bonding character of the surface is actuated,
this phenomenon can lead to both lifelike ``inchworm'' behavior
of the filaments and ensemble movement. We illustrate, through simulation,
how pattern formation may be achieved through this mechanism.
\end{abstract}
\maketitle

\section{introduction}

In previous work[1], we explored a toy model
for the Amyloid beta fibril CF-PT and a class of conformations for
that model which lead to large bonding energies when in contact with
a flat surface. Motivated by the proposition that CF-PT (cylindrical
filament with periodic thinning) is a precursor filament for PHF (paired
helical filament)[2], we showed that one such
strongly bonded conformation for our model is a helix like that of
PHF. This earlier investigation was limited to a surface with a static,
uniform bonding energy. Here we consider the possibility of a flat,
nonbonding surface with a moving bonding region and discuss the dynamic
conformational change and translation of the filament associated with
moving the region beneath it. To clarify, we do not consider the case
where the surface itself is moving but rather that the region of the
surface which has the potential to bond with the filament varies in
time (due to electrical charge, chemical variation, etc.). As in the
previous study, we consider a ``close contact'' approximation for
the nature of the bonding between the surface and the filament; that
is, the bonding energy between a point on the filament and a point
on the surface is nonzero only if these two points are in contact.

Our model filament consists of a cylinder with a helical bonding pattern,
of period $L$, such that only the patterned region of the filament
may bond with the surface. We will refer to a segment of the filament
of length $L$ where the bonding pattern is in contact with the surface
at the beginning and end of the segment as a ``monomer''. A cartoon
of two monomers is shown below in Figure 1 A. To assume the helical
conformation, each monomer twists to align the bonding pattern with
the surface at both ends, and bends in the center. A sketch of a monomer
in the helical conformation is shown below in Figure 1 B. A consequence
of this bending is that the monomer lifts off the surface in the center
and the length of the monomer projected onto the surface decreases
by some amount $\Delta L_{Bond}$. This is illustrated in Figure 1
C, and a cartoon of a filament comprising three monomers assuming
the helical conformation is displayed in Figure 1 D, shown below.

\begin{figure}[H]
\noindent \begin{centering}
\includegraphics[scale=0.45]{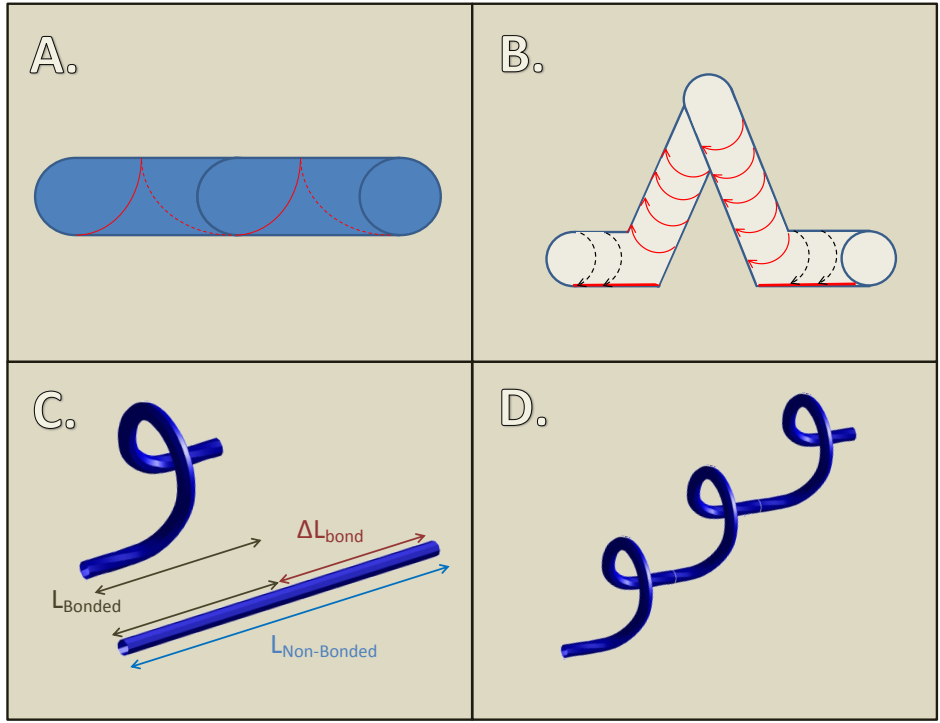}
\par\end{centering}

\caption{A. Cartoon of two filament monomers. B. Sketch of a single monomer
in the helical conformation. The red lines indicate the bonding pattern
and the black dashed lines the twist of the filament. C. Illustration
of the shortened projected length of the helical conformation. D.
Cartoon of a filament comprising three monomers assuming the helical
conformation.}

\end{figure}

\section{Simulation}

With the conformational change discussed above in the presence of
a bonding surface, we know that if we can actuate the surface in such
a way that the bonding character changes with time, we can control
the filament shape as well. Furthermore, we can show that with the
right actuation, the filament is subject to not only a temporary shape
change while bonded, but a net displacement. Thus with continued surface
actuation, filament motility and ``migration'' can be achieved.

This displacement stems from an asymmetric shortening and lengthening
of the filament when a bonding region is introduced and removed. When
the filament binds to the surface, each monomer curls up, as shown
above, decreasing the contact length by an amount $\Delta L_{bond}$
and moving the endpoints of the whole filament inwards. Similarly,
when the filament unbinds, each monomer stretches out, moving the
endpoints outwards. If the bonding region is brought in contact with
and removed from the entire filament simultaneously, this movement
of the endpoints is symmetric and the filament faces no net displacement;
however, if a bonding surface is present beneath one end of the filament
and not the other, the motion of the bonded end can be expected to
be more restricted than that of the unbonded end (due to increased
friction with the surface etc.). This leads to filament motion contrary
to the motion of the bonding region. As the bonding region is introduced,
the filament preferentially shortens from the unbonded end moving
towards the oncoming bonding section. As the bonding region is removed,
the filament preferentially lengthens away from its direction of retraction.
This motion qualitatively resembles that of an inchworm.

When simulating motion due to surface actuation, we need to approximate
how motion of bonded monomers is restricted compared to that of non-bonded
monomers. In reality this will widely vary based on both the filament
and surface materials but for the purposes of constructing an illustrative
simulation, we will assume the following conditions. 1) each bound
monomer confers the same restriction to motion when bound and 2) in
the limit where only one monomer is unbound, only the unbound monomer
moves. Now let $N$ be the total number of monomers in the filament,
$n$ be the number of those bonded, for illustrative purposes assume
a large $\Delta L_{Bond}$ of $\frac{L}{2}$, and denote $\mathbf{p}(1)$
and $\mathbf{p}(N)$ as the positions of the first and last monomers
respectively. We may now consider the displacement of the monomer
at \textbf{$\mathbf{p}(1)$} in the direction of \textbf{$\mathbf{p}(N)-\mathbf{p}(1)$}.

\begin{equation}
Displacement\, of\,\mathbf{p}(1)=\begin{cases}
\begin{array}{c}
contracting\\
\\
\\
extending
\end{array} & \begin{array}{c}
\begin{cases}
\begin{array}{c}
only\,\mathbf{p}(1)\, is\, bound\\
only\,\mathbf{p}(N)\, is\, bound\\
else
\end{array} & \begin{array}{c}
\frac{L}{4}\left(1-\frac{n}{N}\right)\\
\frac{L}{4}\left(1+\frac{n}{N}\right)\\
\frac{L}{4}
\end{array}\end{cases}\\
\begin{cases}
\begin{array}{c}
only\,\mathbf{p}(1)\, is\, bound\\
only\,\mathbf{p}(N)\, is\, bound\\
else
\end{array} & \begin{array}{c}
-\frac{L}{4}\left(1-\frac{n}{N}\right)\\
-\frac{L}{4}\left(1+\frac{n}{N}\right)\\
-\frac{L}{4}
\end{array}\end{cases}
\end{array}\end{cases}
\end{equation}
Using this displacement rule, we may propagate the position of the
filament as we move the bonding region. Motion of a single filament
with a rightward moving, green bonding region is shown below in Figure
2.

\begin{figure}[H]
\noindent \begin{centering}
\includegraphics[scale=0.6]{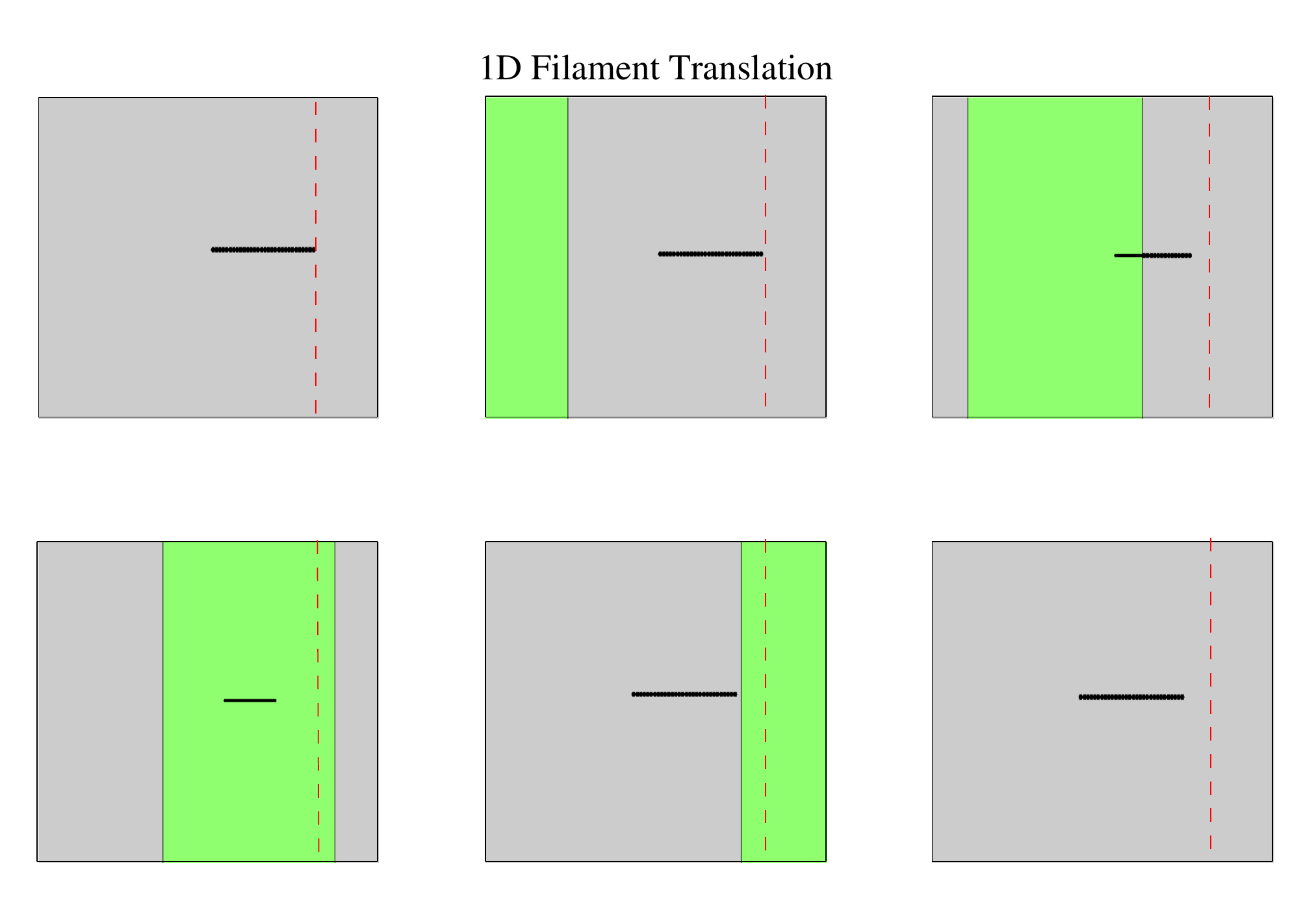}
\par\end{centering}

\caption{Leftward filament migration (time progresses from left to right and
top to bottom) in response to a rightward moving, green bonding section.}
\end{figure}

\medskip{}
We may note that filaments will self-order by length (see Figure 3,
Sub-Figure a, below) as with each pass of the bonding region, a longer
filament experiences a greater translation (due to a larger number
of monomers).

Filament translation in 2D can be achieved by introducing motion of
the bonding region along two perpendicular axes. In this case, the
shift of a filament caused by motion along one axis can oppose that
from the other in part or in full resulting in reduced or null motion
for filaments of the proper orientation. In the simulation displayed
in Figure 3, Sub-Figure b, the bonding region moves from left to right
and bottom to top. If a filament is oriented such that the angle its
tangent axis makes with the $x$-axis is $\frac{3\pi}{4}$ or $\frac{7\pi}{4}$,
the motion resultant from the vertical bonding actuation completely
cancels that from the horizontal and the filament does not move.
Orientations corresponding to angles from $\frac{\pi}{2}$ to $\pi$
and $\frac{3\pi}{2}$ to $2\pi$ experience a similar reduction in
motion.

\medskip{}

\begin{figure}[H]
\subfloat[Self ordering of filaments by length: as time progresses from left
to right, longer filaments move farther.]{\noindent \begin{centering}
\includegraphics[scale=0.35]{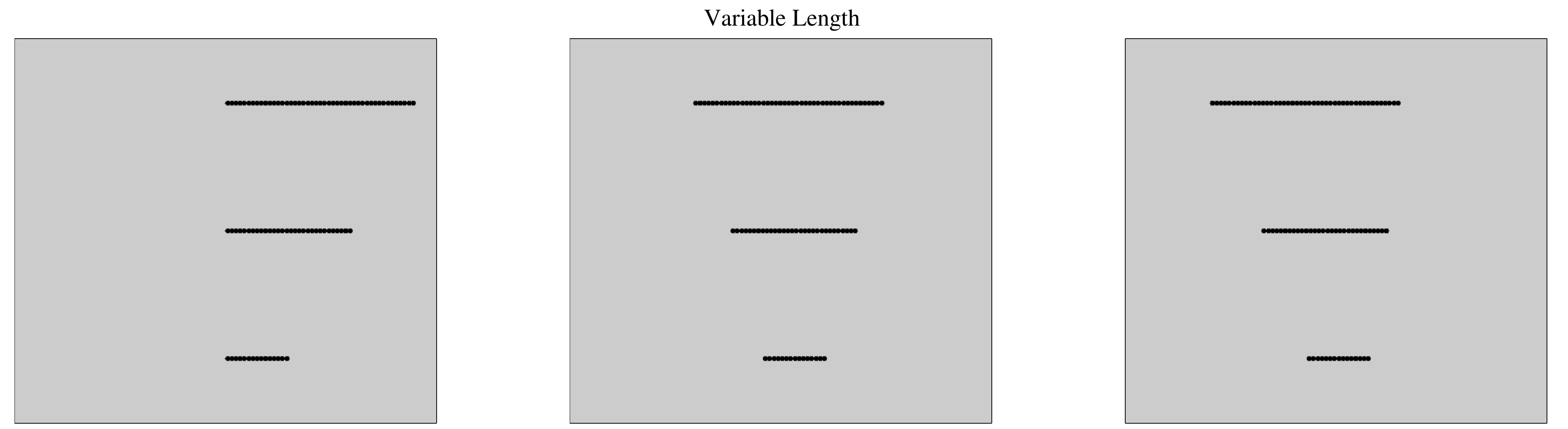}
\par\end{centering}

}

\subfloat[2D translation (Time progresses from left to right.): the bonding
region moves from left to right and bottom to top. If a filament is
oriented such that the angle its tangent axis makes with the $x$-axis
is $\frac{3\pi}{4}$ or $\frac{7\pi}{4}$, the motion resultant from
the vertical bonding actuation completely cancels that from the lateral
and the filament does not move (top left filament). To contrast, a
filament oriented such that the angle its tangent axis makes with
the $x$-axis is $\frac{5\pi}{4}$ or $\frac{9\pi}{4}$ benefits from
maximum translation (bottom left filament). Filaments aligned with
either axis move only along that axis.]{\noindent \begin{centering}
\includegraphics[scale=0.35]{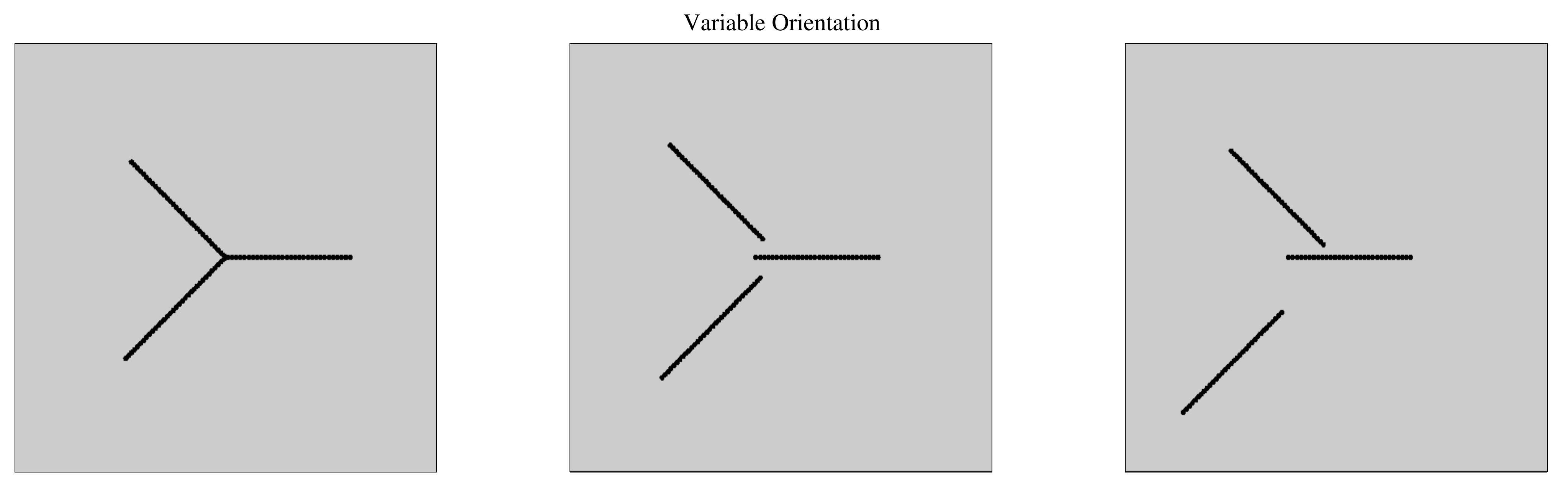}
\par\end{centering}

}

\caption{Self organization of filaments by length and 2D translation.}

\end{figure}

\medskip{}
When many filaments with different orientations are introduced to
a bonding region which moves along two perpendicular axes, as described
above, ring formation occurs. Filaments ``migrate'' with a direction
and magnitude dependent on the angle their tangent axes make with
the $x$-axis.

\begin{figure}[H]
\noindent \begin{centering}
\includegraphics[scale=0.4]{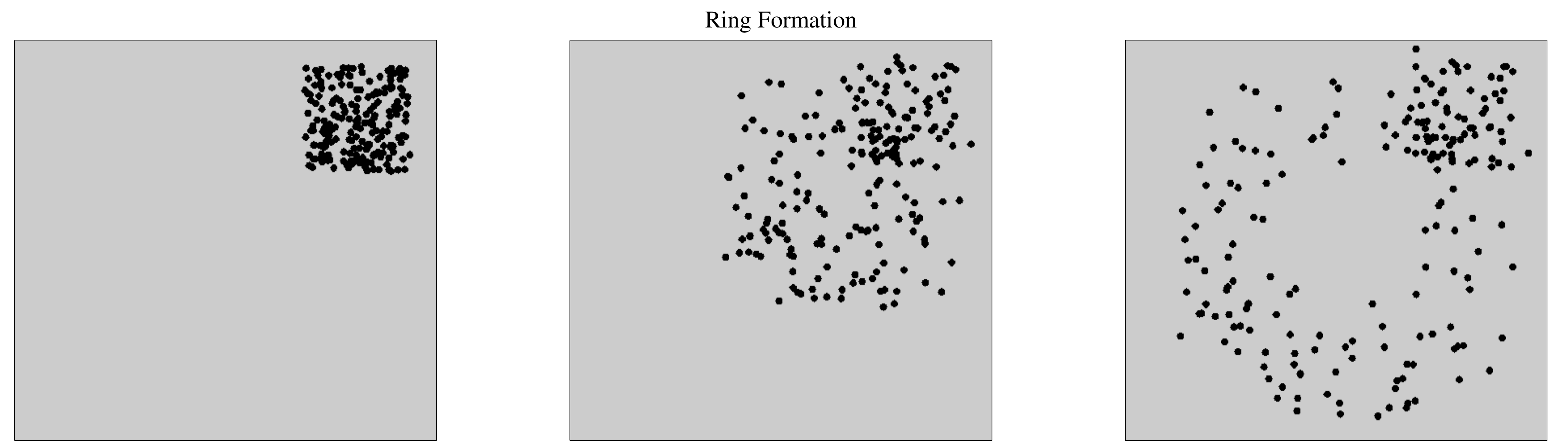}
\par\end{centering}

\caption{2D bonding actuation applied to two-hundred filaments, thirty monomers
in length of variable orientation.}
\end{figure}

\section{discussion}

We've shown that when in contact with a time-varying attractive surface,
helically-patterned filaments may self-sort by length and orientation.
When many of these filaments are placed on such a surface, they can
self-assemble into rings. We hope that this mechanism has potential
applications in the self-organization of materials. Additionally,
the phenomena discussed in this work may have some pathological relevance
in explaining the aggregation of proteins. Many diseases are characterized
by protein disorganization and aggregation e.g. Amyotrophic Lateral
Sclerosis[3] and other neurodegenerative
pathologies[4] though this mechanism is unlikely
to have any physiological relevance, perhaps the idea of surface-mediated
filament migration has some merit. The filament migration described
above, if observed in a bounded region, would result in aggregation
at the boundaries and stochastic surfaces charges seen in biological
systems may resemble the surface actuation described above.

\section{references}

\end{document}